\documentclass[prl,aps,twocolumn,showpacs]{revtex4}

\usepackage{ifpdf}
\ifpdf
\usepackage{subfigure}
\usepackage[pdftex]{graphicx}
\else
\usepackage{graphicx}
\fi

\usepackage{setspace}
\usepackage{psfrag}
\usepackage{epstopdf}
\usepackage{graphicx}
\DeclareGraphicsExtensions{.eps}

\usepackage{amsmath,graphicx,epsfig,bm,color}
\usepackage{amssymb,amsfonts}
\usepackage{rotating}
\usepackage{graphics}
\usepackage{setspace}

\renewcommand{\a}{\alpha}

\renewcommand{\O}{\Omega}

\def\beq{\begin{equation}}
\def\eeq{\end{equation}}

\newcommand\eps{\varepsilon}
\renewcommand{\epsilon}{\varepsilon}

\def\ep{\varepsilon}
\def\lep{{|\mathrm{log }\ \ep|}}
\def\ln{{\mathrm{log }}}

\def\O{\Omega}

\def\a0{\alpha_0}
\def\a{\alpha}

\def\r0{\rho_{0}}

\def\lep{{|\log \ep|}}

\begin{document}
\title{\bf Vortex-peak interaction and lattice shape in rotating two-component Bose-Einstein condensates}
\author{Amandine Aftalion${}^{1}$, Peter Mason${}^{1}$, Juncheng Wei${}^{2}$}
\affiliation {${}^1$CNRS \& Universit\'e Versailles-Saint-Quentin-en-Yvelines,
Laboratoire de Math\'ematiques de Versailles, CNRS UMR 8100, 45
avenue des \'Etats-Unis, 78035 Versailles C\'edex, France \\
${}^2$Department of Mathematics,
 The Chinese University of Hong Kong, Shatin, N.T., Hong Kong.}
\date{\today}
\begin{abstract} When a two-component Bose-Einstein condensate is placed into rotation,
 a lattice of vortices and cores appear. The geometry of this lattice (triangular
 or square) varies according to the rotational value and the intercomponent coupling strengths. In this paper, assuming a Thomas-Fermi regime, we derive a point energy which allows us to determine for which values of the parameters, the lattice goes from triangular to square. It turns out that the separating curve in the phase diagram agrees fully with the complete numerical simulations of the Gross-Pitaevskii equations. We also derive a formula for the
  critical velocity of appearance of the first vortex and prove that the first vortex always appears first in the component with largest
  support in the case of two disks, and give a criterion in the case of disk and annulus.
\end{abstract}
\maketitle

\section{Introduction} When a two-component condensate is set into rotation, topological defects of both order parameters are created, which lead to more exotic defects than in a single component condensate.
 Experiments on two component condensates
 have shown how the condensates can exhibit either triangular or square vortex lattices \cite{sch}.
 According to the values of the interaction strengths, the defect patterns can vary a lot, as illustrated in the numerical simulations \cite{ktu,AM}.
 One specific feature is the appearance of coreless vortices: the existence of a vortex in component-1 corresponds to a peak in component-2 and vice-versa. The interaction between vortices and peaks leads to changes in the geometry of the vortex lattice. We are interested in determining the equations governing this vortex peak behaviour in the Thomas-Fermi regime and estimating the interaction energy between the lattices of the two components.
 Indeed, for a single condensate, the vortex lattice is triangular, while for a two component, the vortex-peak interaction can lead to a square lattice.
  In \cite{tsu}, an asymptotic interaction between two half-quantized vortices is derived for two-component homogeneous condensates. In this paper, we
  want to take into account the nonhomogeneity of the condensate due to the trapping potential and estimate
  the vortex-peak energy according to the parameters of the system.  We derive an energy depending on the location of vortices and peaks and determine for which values
  of the experimental parameters, the lattice goes from triangular to square. These critical values agree well with the ones found from the numerical computations of the full
   Gross-Pitaevskii equations of \cite{AM}. We note that in the rapid rotation regime, using the lowest Landau level approximation, several papers \cite{mh,ko} (see \cite{ktu,C} for a review) have analyzed the transition between triangular to square lattices.  We point out that \cite{barnett} found a point energy with an interaction term $e^{-|p_i-q_j|^2}$, which is different from ours and from \cite{tsu}. We first review relevant results for a single condensate (see \cite{AD}), before moving to the derivation of
   homogeneous equations and the computation of the interaction term in two-component condensates.

 For a single component condensate, the wave function minimizes the energy
 \beq E_{g,\O}(\psi)=\int \frac 12 |\nabla \psi-i\O\times r \psi|^2+\frac 12(V(r)-\O^2r^2)|\psi|^2
  +\frac g2 |\psi|^4\eeq under $\int |\psi|^2=1$, where $\Omega =\Omega e_z$ is the rotation, $V(r)$ is the trapping potential and in most cases $V(r)=r^2$. We will denote by $\nabla_\O$ the operator $\nabla -i\Omega\times r$.
 For $g$ large, at $\Omega=0$, the ground state $\eta$ of $E_{g,0}$ approaches  the inverted parabola $$\frac 1{2g} (\lambda -r^2)$$ in the disk of radius $R^2=\lambda=2\sqrt {g/\pi}$, and goes to 0 outside the disk.  If the problem is rescaled on a disk of size 1, then the analysis of the vortex cores leads to a vortex of size $1/\sqrt g$ and, close to the core, the wave function behaves like $f(r)e^{i\theta}$ where
    $f$ is the solution tending to 1 at infinity of
    \beq \label{f1comp}f''+\frac {f'}r-\frac f{r^2}+f(1-f^2)=0.\eeq This is
    the equation of a vortex core in a uniform system. In the case of a single condensate, from the equation of the vortex core, one can estimate the energy of vortices, the critical velocity for the nucleation
    of the first vortex and the interaction energy between vortices  \cite{AD,F,CD,A} which is
    \beq\label{int1comp}-\sum_{i\neq j} \ln |p_i-p_j|+\sum_i|p_i|^2\eeq where $p_i$ are the location of the vortex cores. Numerically, the minimization of (\ref{int1comp}) yields an almost triangular lattice for a large number of vortex points.

    The aim of this paper is to describe the equivalent of (\ref{f1comp})-(\ref{int1comp})
    in the case of two-component condensates. We define $g_i$ to be the intra-component coupling strength for component $i$ and $g_{12}$ to be the inter-component coupling strength.
     For simplicity, we assume equal masses for the atoms in each component and equal trapping
     potentials, but a general case could be handled. The ground state of a two-component condensate is given by the infimum of
\begin{equation}\label{GS}
	\begin{split} E_{g_1,g_2,g_{12},\Omega}(\psi_1,\psi_2)=E_{g_1,\Omega}(\psi_1)+E_{g_2,\Omega}(\psi_2)\\+ {g_{12}}\int |\psi_1|^2 |\psi_2|^2
\end{split}\eeq
under $\int |\psi_1|^2=N_1$,
    $\int |\psi_2|^2=N_2$.
We  set $g_1=\a_1 g$, $g_2=\a_2 g$, $g_{12}=\a_0 g$
 where $g$ is large, so that $\eps=1/\sqrt g$ is small. We change
  wave functions to $\psi_1(x,y)=\sqrt \eps u_1(x\sqrt \eps,y \sqrt \eps)$,
  $\psi_2(x,y)=\sqrt \eps u_2(x\sqrt \eps,y\sqrt \eps)$. Calling $\alpha=(\a_0,\a_1,\a_2)$,
   the energy we want to minimize is
   \begin{equation}\begin{split}\label{E2comp}E_{\a,\O}(u_1,u_2)=\int &\frac {\ep^2}2 |\nabla u_1|^2
    +\frac 12 r^2 |u_1|^2 + \frac{\a_1}2 |u_1|^4\\&-\eps\Omega\times r(i u_1,\nabla u_1)\\
    &+\frac {\ep^2}2 |\nabla u_2|^2
    +\frac 12 r^2 |u_2|^2 +\frac {\a_2}2 |u_2|^4\\&-\eps\Omega\times r(i u_2,\nabla u_2)\\
	&+{\a_0} |u_1|^2 |u_2|^2
    \end{split}
\end{equation}
 where $(iu,\nabla u)=iu\nabla \bar u-i\bar u\nabla u$.
 For $\O=0$, the ground state is real valued and we denote it by $(\eta_1,\eta_2)$. It is a solution of \beq\label{eta1}-\eps^2\Delta \eta_1+r^2\eta_1+2\a_1\eta_1^3+2\a_0\eta_2^2\eta_1=\mu_1\eta_1\eeq \beq\label{eta2}
-\eps^2\Delta \eta_2+r^2\eta_2+2\a_2\eta_2^3+2\a_0\eta_1^2\eta_2=\mu_2\eta_2.\eeq The shape of the ground state
 varies according to $\alpha$ and when $\a_0^2-\a_1\a_2\leq 0$, can be either two disks or a disk and an annulus, as we will see below.

\section{Reduction to the core equations}

We recall that $(\eta_1,\eta_2)$ is the ground state for  $\Omega=0$ and we consider $(u_1,u_2)$ a ground state of $E_{\a,\O}$. We call $(f_1,f_2)$ such that $ u_1= \eta_1 f_1$ and $u_2= \eta_2 f_2$. We expect $\eta_i$
 to include the slow varying profile and $f_i$ to include the vortex
 or peak contribution, so that $f_i$ is $1$ almost everywhere
 except close to the vortex and peak cores. We want to write
  the energy of $(u_1,u_2)$ as the energy of $(\eta_1,\eta_2)$ plus a rest, which is the energy that we are going to study. This follows a trick introduced in \cite{LM}, and used for single Bose Einstein condensates in \cite{AD,A}. We multiply (\ref{eta1}) by $\eta_1 (|f_1|^2-1)$ and (\ref{eta2}) by
 $\eta_2 (|f_2|^2-1)$, and integrate and add the two equations, which yields the identity
  \begin{multline}\int \frac {\ep^2}2 |\nabla \eta_1|^2 (|f_1|^2-1)+\ep^2\eta_1f_1\nabla \eta_1\cdot \nabla f_1\\+\frac 12 r^2\eta_1^2(|f_1|^2-1)+\a_1 \eta_1^4 (|f_1|^2-1)+\a_0\eta_2^2\eta_1^2 (|f_1|^2-1)\\
  + \frac {\ep^2}2 |\nabla \eta_2|^2 (|f_2|^2-1)+\ep^2\eta_2f_2\nabla \eta_2\cdot\nabla f_2\\+\frac 12 r^2\eta_2^2(|f_2|^2-1)+\a_2 \eta_2^4 (|f_2|^2-1)+\a_0\eta_2^2\eta_1^2 (|f_2|^2-1)=0.\label{id} \end{multline} Note that the Lagrange multiplier term has disappeared because $u_i$ and $\eta_i$ are normalized similarly.  We replace $(u_1,u_2)$ by $(f_1\eta_1,f_2\eta_2)$ into the energy (\ref{E2comp}),
  use the identity (\ref{id}) and find
  \begin{multline}E_{\a,\O}(u_1,u_2)=E_{\a,0}(\eta_1,\eta_2)+F_{\a,\O}(f_1,f_2)\hbox{ where }\\ F_{\a,\O}(f_1,f_2)=\int \frac {\ep^2}2 \eta_1^2 |\nabla f_1|^2
  -\eps \eta_1^2 \Omega\times r(i f_1,\nabla f_1)\\+\frac 12 \a_1 \eta_1^4 (|f_1|^2-1)^2
+  \a_0 \eta_1^2 \eta_2^2 (1-|f_1|^2) (1-|f_2|^2)  \\ +\frac {\ep^2}2 \eta_2^2 |\nabla f_2|^2
  -\eps \eta_2^2 \Omega\times r(i f_2,\nabla f_2)+\frac 12 \a_2 \eta_2^4 (|f_2|^2-1)^2.
    \label{split}\end{multline}
  This splitting of energy does not assume anything about the scales of energy: it is an exact identity.
  We point out that as soon as $\a_1\a_2-\a_0^2\geq 0$, then the quadratic form in the energy $F_{\a,\O}$
   is positive and minimizing $E_{\a,\O}$ in $(u_1,u_2)$ amounts to minimizing $F_{\a,\O}$
    in $(f_1,f_2)$.

   Now we assume that  we scale everything close to a point $p$ where $\eta_1^2=\rho_1$,
    $\eta_2^2=\rho_2$, and $f_1, f_2$ can be written as functions of $p+|r-p|/\eps$.
%and we call $\Omega_\eps=\eps\O$.
Then, in the new variable $\tilde r=|r-p|/\eps$, the functions $f_1, f_2$ are a ground state of \begin{multline} {\mathcal F}_{\a,\O}(f_1,f_2)=\int \frac 12 \rho_1 |\nabla f_1|^2
  -\ep \rho_1 \Omega\times r(i f_1,\nabla f_1)\\+\frac 12 \a_1 \rho_1^2 (|f_1|^2-1)^2
+  \a_0 \rho_1 \rho_2 (1-|f_1|^2) (1-|f_2|^2)  \\ +\frac 12 \rho_2 |\nabla f_2|^2
  -\ep \rho_2 \Omega\times r(i f_2,\nabla f_2)+\frac 12 \a_2 \rho_2^2 (|f_2|^2-1)^2
    \label{fi}\end{multline}and solve the system
 \begin{multline}-\rho_1\Delta f_1-i \ep\Omega\times r \rho_1 \nabla f_1+2\a_1 \rho_1^2(|f_1|^2-1) f_1\\+2\a_0\rho_1\rho_2 f_1 (|f_2|^2-1)=\tilde \lambda_1 f_1\\
 -\rho_2\Delta f_2-i \ep\Omega\times r \rho_2\nabla f_2+2\a_2 \rho_2^2(|f_2|^2-1) f_2\\+2\a_0 \rho_1\rho_2 f_2 (|f_1|^2-1)=\tilde \lambda_2 f_2.
 \end{multline}
This is exactly the system studied in \cite{tsu,LW} for a homogeneous condensate. The splitting of energy has allowed us to reach a homogeneous system.
 Assuming a vortex in component-1 and a spike in component-2, we have $f_1=v_1(r) e^{i\theta}$ and $f_2=v_2 (r)$.
%which yields
%\begin{eqnarray}\label{vs1}- v_1''-\frac {v_1'}r+\frac {v_1} {r^2}- \O_\eps  v_1+2\a_1 \rho_1 (v_1^2-1) v_1 +2\a_0\rho_2 v_1 (v_2^2-1)= \lambda_1 v_1\\
% -v_2''-\frac {v_2'}r+2\a_2 \rho_2 (v_2^2-1) v_2 +2\a_0 \rho_1 v_2 (v_1^2-1)= \lambda_2 v_2.\label{vs2}
% \end{eqnarray}
 We expect that $v_1, v_2$ tend to 1 at infinity so that  $\tilde \lambda_1=\ep\Omega$ and $\tilde \lambda_2=0$.
  This yields the following system
\begin{eqnarray}\label{eqvs1}-\frac {(rv_1')'}r+\frac {v_1} {r^2}+2\a_1 \rho_1 (v_1^2-1) v_1 +2\a_0\rho_2 v_1 (v_2^2-1)= 0\ \ \\
 -\frac {(rv_2')'}r+2\a_2 \rho_2 (v_2^2-1) v_2 +2\a_0 \rho_1 v_2 (v_1^2-1)= 0.\ \ \label{eqvs2}
 \end{eqnarray}
  From this system, asymptotic expansions can be obtained for $v_1$ and $v_2$ at infinity: $v_1 (r)-1\sim -\gamma_1/r^2$ and $v_2(r)-1\sim \gamma_2/r^2$ for some constants $\gamma_1$ and $\gamma_2$.  Equations (\ref{eqvs1})-(\ref{eqvs2}) at infinity imply that  $\a_2\rho_2\gamma_2=\a_0\rho_1\gamma_1$ and $1-4\a_1\rho_1\gamma_1+4 \a_0\rho_2\gamma_2=0$, thus \beq\label{gamma}\gamma_1=\frac 1 {4\rho_1\a_1 \Gamma_{12}}\hbox{ and }\gamma_2=\frac {\a_0} {4\rho_2\a_1\a_2 \Gamma_{12}}\eeq
 where \begin{equation}\label{gamma12}\Gamma_{12}=1-\frac{\a_0^2}{\a_1\a_2}.\end{equation}In particular, \begin{equation}\a_0\gamma_1\gamma_1 \rho_1\rho_2=\frac{1-\Gamma_{12}}{16 \a_1\Gamma_{12}^2}.\label{agam}\end{equation}
In order to fully analyze the system (\ref{eqvs1})-(\ref{eqvs2}), we need to have information
 on $\rho_1$, $\rho_2$, that is the ground states $\eta_1$, $\eta_2$:  in particular, we need to know whether
 the supports of $\eta_1$, $\eta_2$ are disks or annuli, and where they reach their maximum.
%%%%%%%%%%%%%%%%%%%%%%%%%%%%%%%%%%%%%%%%%%%%%%%%%%%%%%%%%%%%%%%%%%
%%%%%%%%%%%%%%%%%%%%%%%%%%%%%%%%%%%%%%%%%%%%%%%%%%%%%%%%%%%%%%%%%%%%

\section{Thomas-Fermi profile of the ground state}
We recall some  properties of the solutions of (\ref{eta1})-(\ref{eta2}) obtained in \cite{AM}.
 The following non-dimensional parameters are introduced
\begin{eqnarray}
    \label{con}
\Gamma_1&=&1-\frac{\a_0}{\a_1}\\
\Gamma_2&=&1-\frac{\a_0}{\a_2}.
\end{eqnarray}
%\subsection{Two disks}
To begin, assume that both components are circular with radii $R_1$ and $R_2$ and with $R_1<R_2$.  When $\a_1\a_2-\a_0^2\geq 0$, that is
 $\Gamma_{12}\geq 0$, and $\ep$ is small, both components are in the Thomas-Fermi (TF) regime,  and the density profiles for $r<R_1$ are
\begin{eqnarray}
\label{tff}
|\eta_1|^2&=&\frac{1}{2\a_1\Gamma_{12}}\left(\mu_1-\frac{\a_0}{\a_2}\mu_2
-r^2\Gamma_2\right)\\
\label{tff2}
|\eta_2|^2&=&\frac{1}{2\a_2\Gamma_{12}}\left(\mu_2-\frac{\a_0}{\a_1}\mu_1-
r^2\Gamma_1\right)
\end{eqnarray}
and for $R_1<r<R_2$ are
\begin{equation}
\label{tff3}
|\eta_2|^2=\frac{\mu_2-r^2}{2\a_2}
\end{equation}
with $|\eta_1|^2=0$.
The chemical potentials $\mu_1$ and $\mu_2$, and the radii, $R_1$ and $R_2$, are to be found. In addition we have the normalisation condition
\begin{equation}
\label{norm1}
\int|\eta_k|^2d =N_k,
\end{equation}
where, for generality, $N_1\neq N_2$. We denote $\tilde \a_k=N_k \a_k$ and $\tilde\a_0=\sqrt{N_1N_2}\a_0$ and get
\begin{eqnarray}
\label{eq:r1}
R_1&=&\left(\frac{4\tilde{\a}_1\Gamma_{12}}{\pi\Gamma_2}\right)^{1/4},\\
R_2&=&\left(\frac{4(\tilde{\a}_2+\tilde{\a}_1(1-\Gamma_1))}{\pi}\right)^{1/4},\label{eq:r2} \\
\nonumber
\mu_1&=&\left(\frac{4\tilde{\a}_1\Gamma_{12}\Gamma_2}{\pi}\right)^{1/2}\\ &&+
(1-\Gamma_2)\left(\frac{4}{\pi}\left[\tilde{\a}_2+\tilde{\a}_1(1-\Gamma_1)\right]\right)^{1/2}\label{mu1}
,\\
\label{mu2}
\mu_2&=&\left(\frac{4}{\pi}\left[\tilde{\a}_2+\tilde{\a}_1(1-\Gamma_1)\right]\right)^{1/2}.
\end{eqnarray}

We find from (\ref{tff}) and (\ref{gamma}) that \beq\label{rho1eq}\rho_1=\eta_1^2 (0)=\frac{\Gamma_2R_1^2}{2\a_1\Gamma_{12}}=
\sqrt{ \frac{\Gamma_2N_1}{\pi\a_1\Gamma_{12}}}\eeq  and $$\gamma_1=\sqrt{\frac \pi{16N_1\a_1\Gamma_2\Gamma_{12}}}$$
 while
 \beq\label{rho2eq} \rho_2=\eta_2^2 (0)=\frac {1} {\a_2}\left (
  (\frac 1 \pi (N_2 \a_2+N_1 \a_0))^{1/2}-\a_0 (\frac {N_1\Gamma_2}{\pi\a_1\Gamma_{12}})^{1/2}\right )\eeq and $\gamma_2$ follows from (\ref{gamma}).
% \beq\label{gammai} \gamma_1\gamma_2=
%   \frac{N_1\a_0\Gamma_2}{16\pi\a_1^2\a_2^2\Gamma_{12}^2} \frac {N_2\a_2\Gamma_{12}
%    N_1\a_0\Gamma_1}{\sqrt{N_1\Gamma_2(N_2\a_2+N_1\a_0)}+N_1\a_0\Gamma_1}\eeq

 Equations (\ref{eq:r1})-(\ref{eq:r2}) are valid provided $\Gamma_{12}/\Gamma_2>0$ (to ensure that $R_1$ and $\mu_1$ are real) and $\tilde{\a}_1\Gamma_1<\tilde{\a}_2\Gamma_2$ (to ensure that $R_2>R_1$). If instead the initial assumption on the size of the radii was taken to be $R_1>R_2$, then the appropriate expressions would also be given by Eq.'s (\ref{eq:r1})-(\ref{eq:r2}), however with the indices $1$ and $2$ alternated. In this case, the conditions would be $\tilde{\alpha}_1\Gamma_1>\tilde{\alpha}_2\Gamma_2$ and $\Gamma_{12}/\Gamma_1>0$.

Returning now to $R_2>R_1$, one must ensure that $|\eta_2|^2>0$ for all $r<R_1$ to have a disk rather than an annulus. Suppose  that there is a point at the origin, where $|\eta_2|^2=0$. Then, from Eq. (\ref{tff2}),
\begin{eqnarray}
    \a_0&=&\bar{\a}_{0}=\frac{\a_1\mu_2}{\mu_1}\nonumber\\
    \label{spatsep}
    &=&\frac{N_1\a_1}{2(N_1+N_2)}+\frac{1}{2}\sqrt{\frac{\a_1^2N_1^2}{(N_1+N_2)^2}
    +\frac{4N_2\a_1\a_2}{N_1+N_2}}.%\\
%   &>&\frac{N_1g_1}{N_1+N_2}.
\end{eqnarray}
%Now assume that $\bar{g}_{12}^2/g_1g_2>1$, or equivalently,
%\begin{eqnarray}
%   &&\frac{1}{g_1g_2}\left(\frac{N_1g_1}{N_1+N_2}\right)^2>1\nonumber\\
%   &\Rightarrow& N_1(g_1-g_2)>g_2N_2^2+2g_2N_1N_2
%   \label{df}
%\end{eqnarray}
%which is in contradiction since the left hand side of (\ref{df}) is negative (by assumption $g_2>g_1$) while the right hand side is positive.
The existence of some $\Gamma_{12}$ at which the density in component-2 hits zero at the origin is the indication of a spatial separation of the components.  Notice that this critical value for $\Gamma_{12}$ is independent of $\Omega$. In the spatial separation regime, component-1 is circular while component-2 is annular, provided $R_2>R_1$. It is not possible for an annulus to develop in component-1 if $R_2>R_1$; this can be seen by writing down the TF density expressions for an annular component-1 and a circular component-2 in which the chemical potentials become multi-valued. Similarly, an annulus can only develop in component-1 if $R_1>R_2$. Thus under the assumption $R_2>R_1$, an annulus can only develop in component-2 and
 the condition to have two disks is thus $\a_0<\bar \a_0$.

\section{Vortex interaction}
 Let us call $\rho_{TF,1}$, $\rho_{TF,2}$ the Thomas-Fermi limits of $|\eta_1|^2$ and $|\eta_2|^2$ given by
(\ref{tff})-(\ref{tff2})-(\ref{tff3}) and (\ref{eq:r1}), (\ref{eq:r2}), (\ref{mu1}), (\ref{mu1bis}), (\ref{mu2})
 in the case of two disks. Then\beq \rho_{TF,1}=\frac{\Gamma_2}{2\a_1 \Gamma_{12}} (R_1^2-r^2)\label{rtf1}\eeq
 \beq \rho_{TF,2}=\frac{\Gamma_1}{2\a_2 \Gamma_{12}} (R_1^2-r^2)+\frac 1{2\a_2} (R_2^2-R_1^2) \hbox{ if }r<R_1\label{rtf2}\eeq\beq\hbox{ and }\frac {1}{2\a_2} (R_2^2-r^2) \hbox{ if }r>R_1 .\label{rtf2bis} \eeq We want to estimate the various terms in the energy $F_{\alpha,\Omega}$ as in \cite{AD,A} and we are going to show that, if $p_i$ are the vortices for component-1 and $q_j$ are the vortices for component-2, then they minimize  the point energy
\begin{multline}\label{renorm}
	-\pi \ep^2 \sum_{i\neq j}\rho_1 \log |p_i-p_j|
	-\pi \ep^2 \sum_{i\neq  j}\rho_2 \log |q_i-q_j|\\
	+\pi\left(- \ep^2\lep\frac{\Gamma_2}{2\alpha_1\Gamma_{12}} +\ep\Omega\rho_1\right)\sum_i |p_i|^2\\
	+\pi\left(- \ep^2\lep\frac{\Gamma_1}{2\alpha_2\Gamma_{12}} +\ep\Omega\rho_2\right)\sum_i |q_i|^2\\+\pi \frac {1-\Gamma_{12}}{16\Gamma_{12}^2}\left ( \frac 1 {\a_1}
 + \frac 1 {\a_2}\right )  \ep^4\lep \sum_{i\neq j}  \frac 1 {|p_i-q_j|^2}.
	\end{multline}

\subsection{Estimate of the kinetic energy term}
 Let us call $p_i$ the vortices in component-1, and $q_j$ in component-2.
  Then the kinetic energy term $(1/2)\int \eta_1^2\ep^2 |\nabla f_1|^2$ provides a leading order term due to the kinetic energy of the phase
  (which behaves locally like $1/r$ outside a disk of radius $\ep$ around each vortex), which is
  \beq\label{kin}\pi \ep^2 \sum_i\rho_{TF,1} (p_i) \lep-\pi \ep^2 \sum_{i\neq j}\rho_{TF,1} (p_i)
  \log |p_i-p_j|\eeq with a similar term for component-2, where $p_i$ is replaced by $q_j$.

 \subsection{Rotation term}
  We call $X_1(r)$ the primitive of $-r\rho_{TF,1} (r)$ which vanishes at $R_1$ and
  $X_2(r)$ the primitive of $-r\rho_{TF,2} (r)$ which vanishes at $R_2$. Then for $r<R_1$
  \beq\label{X1}X_1(r)=\frac{\Gamma_2}{8\a_1 \Gamma_{12}} (R_1^2-r^2)^2,\eeq
  \beq\label{X2}X_2(r)=\frac{\Gamma_1}{8\a_2 \Gamma_{12}} (R_1^2-r^2)^2+\frac {1}{8\a_2} (R_2^2-R_1^2)(R_2^2+R_1^2-2r^2).\eeq
  Thus the rotation term $-\ep\Omega \int \eta_1^2\times r(i f_1,\nabla f_1)$ is well
  approximated by $-\ep\Omega \int \nabla X_1\times (i f_1,\nabla f_1)$. An integration by parts around each vortex  yields \beq\label{rot}-2\pi\ep\Omega \sum_i X_1(p_i)\eeq with a similar contribution for component-2.

\subsection{First vortices}
   The leading order approximation of the kinetic and rotation energy
    yields (assuming vortices at points $p_i$ for component-1 and $q_j$ for component-2):
  \begin{multline}\label{renomm}\pi \ep^2 \sum_i\rho_{TF,1} (p_i) \lep+\pi \ep^2 \sum_j\rho_{TF,2} (q_j) \lep\\-2\pi\ep\Omega \sum_i X_1 (p_i)-2\pi\ep\Omega \sum_j X_2 (q_j).\end{multline} The energy is minimized  by a configuration such that
   $\rho_{TF,1}/X_1$ or $\rho_{TF,2}/X_2$
   reaches its minimum. We find that \beq\label{maxXr1}\frac {X_1 (r)} {\rho_{TF,1}(r)}=
   \frac 14(R_1^2-r^2)\eeq \beq\label{maxXr2}\frac {X_2 (r)} {\rho_{TF,2}(r)}=
   \frac 14(R_1^2-r^2)+\frac{1}{4}\frac {(R_2^2-R_1^2)(R_2^2-r^2)}{\frac{\Gamma_1}{\Gamma_{12}} (R_1^2-r^2)+(R_2^2-R_1^2)}\eeq
% and that
% \begin{equation}
% 	\label{maxr1X}
% 	\frac {\rho_{TF,1}(r)}{X_1 (r)}=\frac{4}{(R_1^2-r^2)}
% \end{equation}
% \begin{equation}
% 	\label{maxr2X}
% 	\begin{split}
% 	 \frac{\rho_{TF,2}(r)}{X_2(r)}=&\frac{4}{(R_1^2-r^2)}\\&-\frac{4(R_2^2-R_1^2)(R_2^2-r^2)}{(R_1^2-r^2){\left[\frac{\Gamma_1}{\Gamma_{12}}(R_1^2-r^2)^2+(R_2^2-R_1^2)(R_1^2+R_2^2-2r^2)\right]}}
% \end{split}
% \end{equation}
  This implies that above a critical value $\Omega^c$, vortices become energetically favorable in the system
   and $\Omega^c$ is given from (\ref{renomm}) by
   \beq\label{omegacc}\Omega^c=\frac 12\ep \lep \min_{i, r} \frac {\rho_{TF,i}}{X_i}.\eeq
    For a harmonic potential, and in the case of two disks, the minimum of $\frac {\rho_{TF,i}}{X_i}$ occurs
   at the origin. Furthermore, since
\begin{equation}
\frac{\rho_{TF,2}}{X_2}=\frac{\rho_{TF,1}}{X_1}-D(r)
\end{equation}
where
\begin{equation}
	\begin{split}
	 D(r)=&\frac{4(R_2^2-R_1^2)(R_2^2-r^2)}{(R_1^2-r^2)}\times\\&\frac{1}{\left[\frac{\Gamma_1}{\Gamma_{12}}(R_1^2-r^2)^2+(R_2^2-R_1^2)(R_1^2+R_2^2-2r^2)\right]},
\end{split}
\end{equation}
and given the signs of the parameters, we see that $D(0)>0$ so that we always have ${\rho_{TF,1}}/{X_1}>{\rho_{TF,2}}/{X_2}$.

The first vortex is thus preferred in component-2 (i.e. the component with larger support) and occurs at the origin with the critical velocity given by
\begin{equation}
	\label{omc}
\begin{split}
	 \Omega^c=\frac{\sqrt{\pi}}{\tilde{\a}_2\Gamma_{12}}\ep\lep\Big[\Gamma_{12}&\sqrt{\tilde{\a}_2+\tilde{\a}_1(1-\Gamma_1)}\\&-(1-\Gamma_1)\sqrt{\tilde{\a}_1\Gamma_2\Gamma_{12}}\Big].
\end{split}
\end{equation}
Note that this expression gives $\Omega^c=0$ when $\a_{0}=\bar{\a}_{0}$ (provided $\a_1\neq \a_2$ otherwise $\Omega^c$ reduces to a non-zero constant).
 For some computations in the rest of the paper, we can assume $N_1 \a_1=N_2 \a_2$ so that $R_1=R_2$ and we have a lattice of peaks and vortices close to the origin. Since $\min_{i, r} {\rho_{TF,i}}/{X_i}=4/R_1^2$, we have
    \beq\label{omega}\Omega^c=\ep \lep  \sqrt{\frac{\pi \Gamma_1}{\a_1\Gamma_{12}}}.\eeq

We have plotted $\Omega^c-\Gamma_{12}$ curves for two cases in Fig. \ref{om_crit}, the first with distinct intracomponent coupling strengths and the second with equal intracomponent coupling strengths (where Eq. (\ref{omc}) reduces to Eq. (\ref{omega})) and compared then to the numerical results of \cite{AM} (these parameter sets correspond to sets `ES1' and `ES3' respectively from \cite{AM}). We find good agreement between the two results.

\begin{figure}
\begin{center}
\includegraphics[scale=0.4]{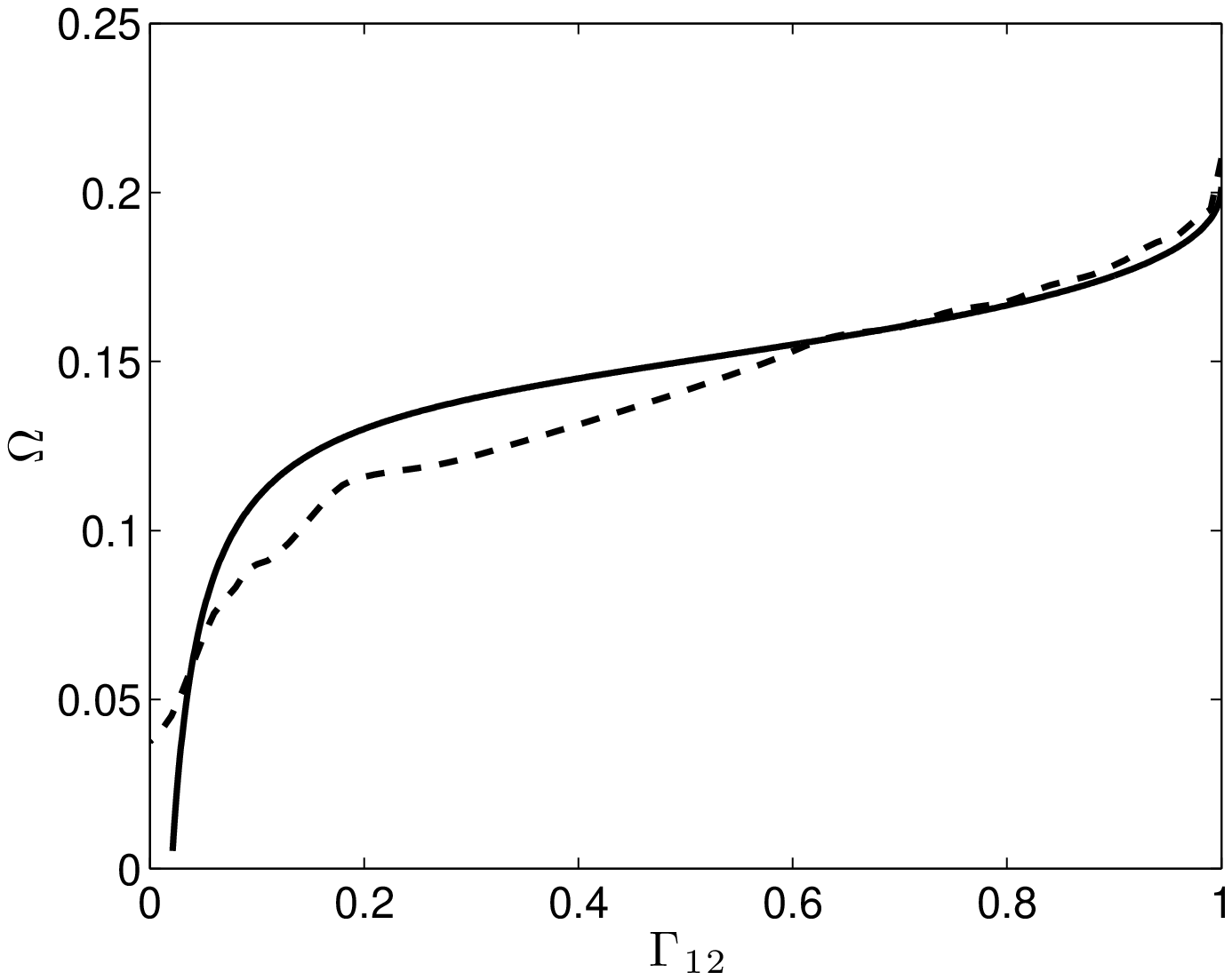}\\
\includegraphics[scale=0.4]{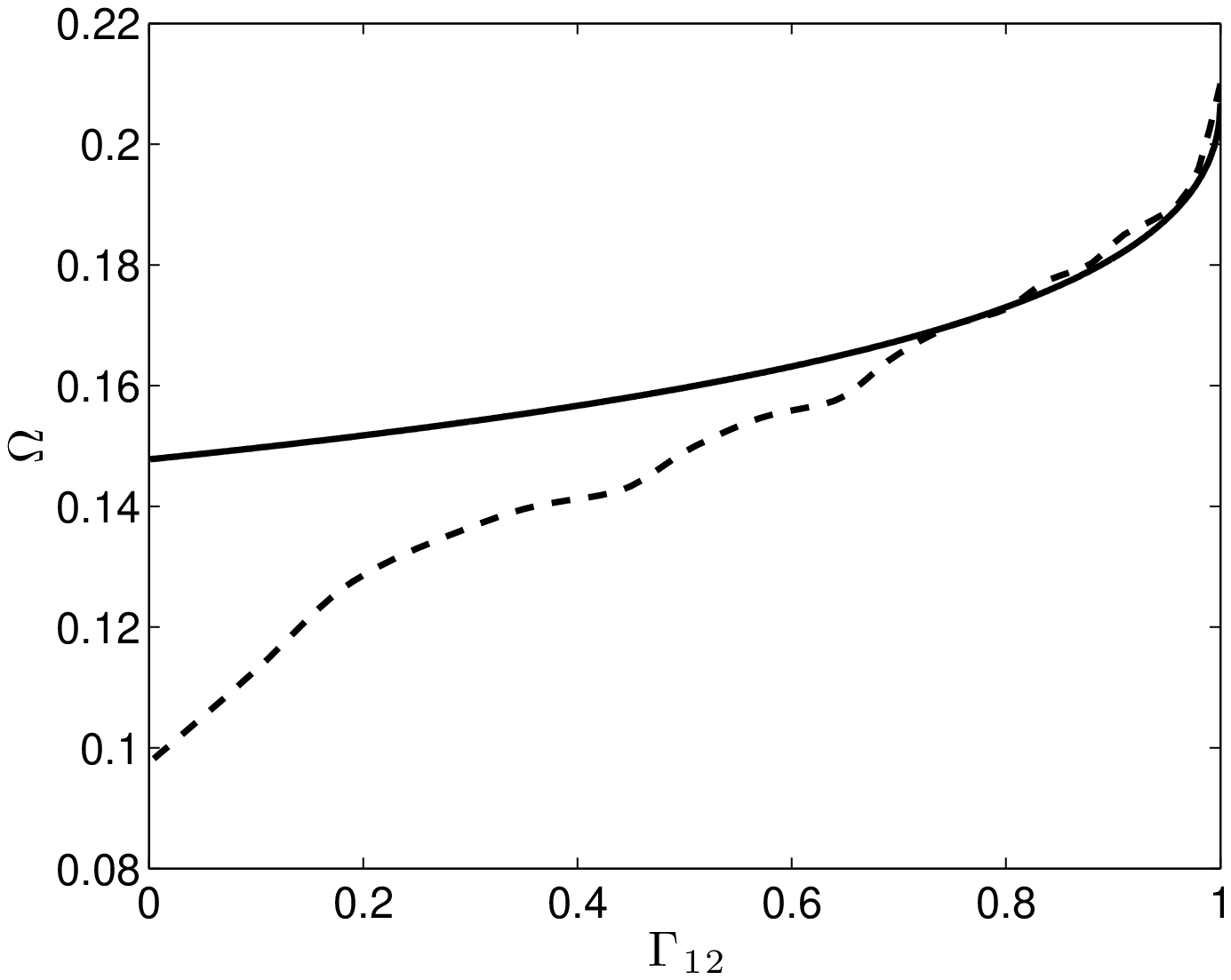}
\end{center}
\caption{The critical velocity for creation of the first vortex plotted analytically from Eq. (\ref{omc}) [solid line] and numerically [dotted line] as a function of $\Gamma_{12}$ for two parameter sets: (a) $\ep=0.0352$, $\a_1=0.97$, $\a_2=1.03$ and (b) $\ep=0.0358$, $\a_1=\a_2=1$.}\label{om_crit}
\end{figure}

\subsection{Energy expansion} We now have to go further into the energy expansion  to estimate the interaction energy.
We assume that the vortices appear close to the origin. Then  (\ref{kin}) and (\ref{rot})
  can be expanded around the origin, using (\ref{rtf1})-(\ref{rtf2bis}) and (\ref{X1})-(\ref{X2}), which yields
  \begin{multline}
	-\pi \ep^2 \sum_{i\neq  j}\rho_1 \log |p_i-p_j|
	-\pi \ep^2 \sum_{i\neq  j}\rho_2 \log |q_i-q_j|\\
	+\pi\ep\left(- \ep\lep\frac{\Gamma_2}{2\alpha_1\Gamma_{12}} +\Omega\rho_1\right)\sum_i |p_i|^2\\
	+\pi\ep\left(- \ep\lep\frac{\Gamma_1}{2\alpha_2\Gamma_{12}} +\Omega\rho_2\right)\sum_i |q_i|^2\label{kinrot}
	\end{multline}

\subsection{Interaction energy}We find from (\ref{split}) that the interaction energy is
 $$\a_0 \rho_1 \rho_2 \int (1-|v_1|^2) (1-|v_2|^2).$$ Near a vortex-peak, this reduces to
 $$\a_0 \rho_1 \rho_2 \gamma_1 \gamma_2 \int \frac 1{r^2_{(1,0)}}
  \frac 1 {r^2_{(0,1)}} $$ where we take the notations of \cite{tsu}: $r_{(1,0)}$
   is the local distance to the vortex in component-1, and $r_{(0,1)}$ is the distance
    to the next peak in component-1, or equivalently to the vortex in component-2.
From (\ref{agam}), we find that the coefficient in front of the integral is equal to
 $ \pi \frac {1-\Gamma_{12}}{16\a_1\Gamma_{12}^2}$. The computations in  \cite{tsu} allow us to estimate the integral  term and we find
  for the interaction term $$\pi \frac {1-\Gamma_{12}}{16\Gamma_{12}^2} \frac 1 {\a_1}
  \ep^4\lep \frac 1{|p_i-q_j|^2}.$$
  This is for a vortex in $v_1$. Of course, if the vortex is in $v_2$, it would be different by a factor $1/\a_2$.

 The interaction energy is thus \beq\label{intere}\pi \frac {1-\Gamma_{12}}{16\Gamma_{12}^2}\left ( \frac 1 {\a_1}
 + \frac 1 {\a_2}\right ) \ep^4\lep\sum_{i\neq j} \frac 1{|p_i-q_j|^2}.\eeq Together with (\ref{kinrot}),
  this leads to (\ref{renorm}).

\section{Numerical simulation of the renormalised energy}

\begin{figure}
\begin{center}
\includegraphics[scale=0.4]{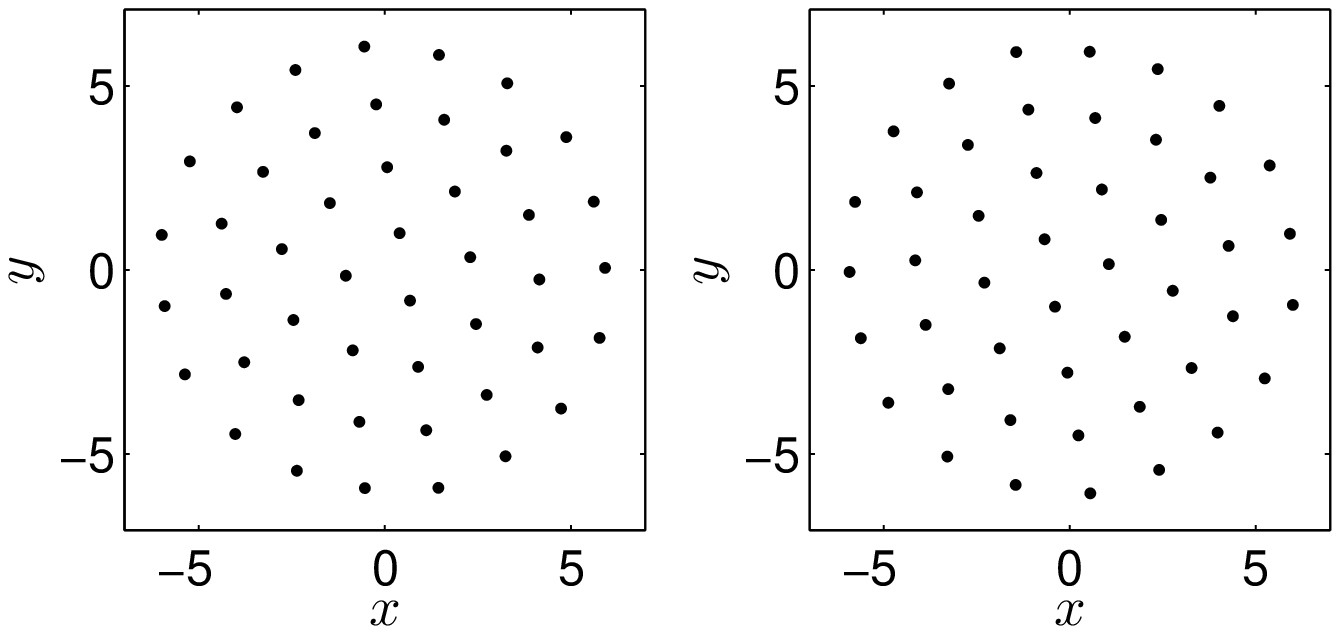}\\
\includegraphics[scale=0.4]{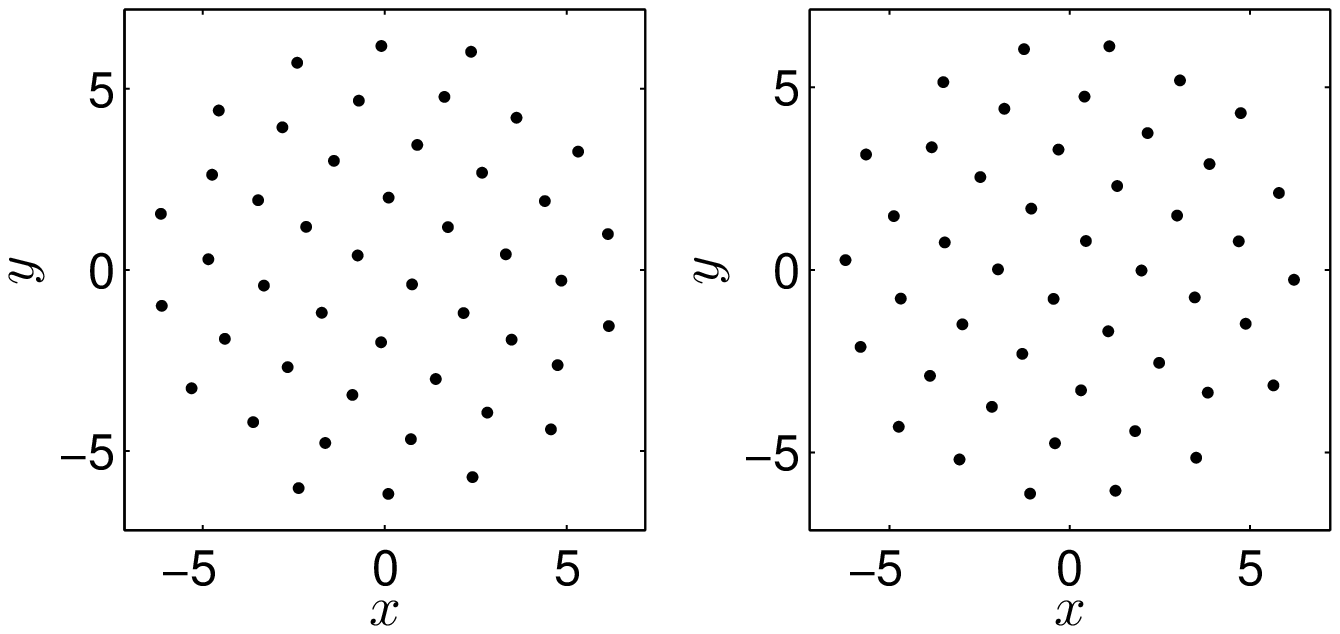}
\end{center}
\caption{ The ground state of the renormalised energy (\ref{renorm1}) calculated for (a) $c_{\omega}=0.05$ and (b) $c_{\omega}=0.3$ with the number of points in each component equal to $N=46$. Component-1 is shown in the left columns and component-2 in the right columns.}\label{latt}
\end{figure}

\begin{figure}
\begin{center}
\includegraphics[scale=0.4]{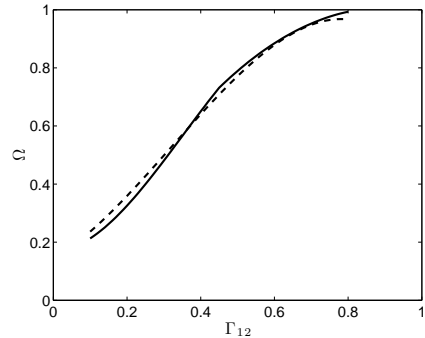}
\end{center}
\caption{ The critical value of $\Omega^{ts}$ as a function of $\Gamma_{12}$ calculated analytically by Eq. (\ref{om_ts}) (solid line) and numerically (dashed line) for the parameters $\ep=0.0358$, $\a_1=\a_2=1$. }\label{latt_num}
\end{figure}

We now want to find the ground state of (\ref{renorm}) when the radii of both components are equal. Under this condition we have $\rho_1=\rho_2$, $\Gamma_1=\Gamma_2$ and $\a_1=\a_2$. This allows us to perform a rescaling that leaves the renormalised energy only dependent on a single parameter. We write $\Omega=\omega\Omega^c$ for $\Omega^c$ defined in (\ref{omega}). Then
\begin{equation}
	\rho_1\left(- \frac{\lep}{R_1^2} +\frac{\Omega}{\ep}\right)=\rho_1\lep\left(\omega\sqrt{\frac{\pi\Gamma_2}{\a_1\Gamma_{12}}}-\frac{1}{R_1^2}\right)
\end{equation}
which implies that we can rescale the $p_i$'s and $q_j$'s as $p_i=\gamma \tilde p_i$ and $q_j=\gamma \tilde q_j$  with
\begin{equation}
	 \gamma^2=\frac{1}{2\lep\left(\omega\sqrt{\frac{\pi\Gamma_2}{\a_1\Gamma_{12}}}-\frac{1}{R_1^2}\right)}
\end{equation}
so that the new energy is
\begin{multline}\label{renorm1}
	\frac{1}{2}\pi\ep^2\rho_1\Bigg[ -\sum_{i \neq j}\log |p_i-p_j|^2
	-\sum_{i\neq  j}\log |q_i-q_j|^2\\
	+\sum_i |p_i|^2
	+\sum_i |q_i|^2+c_{\omega} \sum_{i\neq j}  \frac 1 {|p_i-q_j|^2}\Bigg]
	\end{multline}
with
\begin{equation}
	 c_{\omega}=\frac{\pi(1-\Gamma_{12})}{4\a_1\Gamma_{12}^2}(2\omega-1)\ep^2\lep^2.\label{comega}
\end{equation}

We simulate this renormalised energy (\ref{renorm1}) using a conjugate gradient method varying the parameter $c_{\omega}$ and the number
 of vortex points. For a fixed number of lattice points $N$, when increasing $c_{\omega}$, the ground state lattice goes from triangular to square at a critical $c_{\omega}^{ts}$. Note that when $N$ gets large, $c_{\omega}^{ts}$ no longer depends on $N$. In Fig. \ref{latt} we plot the ground state for two values of $c_{\omega}$ which give a triangular ($c_{\omega}=0.05$) and a square lattice ($c_{\omega}=0.3$) when $N=46$.

For each $N$, we can calculate the critical $c_{\omega}^{ts}$ and compare this with the simulations on the full GP equations, as performed in \cite{AM} (they provide the appropriate value of $N$). From (\ref{comega}), we thus find the critical value of $\Omega$ for which the lattice goes from triangular to
square:
\begin{equation}
	 \Omega^{ts}=\left(\frac{1}{2}+\frac{2c_{\omega}^{ts}\a_1\Gamma_{12}^2}{\pi(1-\Gamma_{12})\ep^2\lep^2}\right)\sqrt{\frac{\pi\Gamma_2}{g_1\Gamma_{12}}}\lep
\end{equation}
It turns out that when $\Omega$ gets close to 1, the condensate expands and one has to include in the TF profile
 a term $(1-\Omega^2)r^2$ instead of just $r^2$. The radii $R_1$ and $R_2$ vary like $(1-\Omega^2)^{1/4}$.
 This changes $c_\omega$ from (\ref{comega}) to \begin{equation}
	c_{\omega}=\frac{\pi(1-\Gamma_{12})}{4\a_1\Gamma_{12}^2}\left(\frac {2\omega-\sqrt{1-\Omega^2}}{\sqrt{1-\Omega^2}}\right )\ep^2\lep^2
\end{equation} so that if we define \begin{equation}
	 \beta_{ts}=\left(\frac{1}{2}+\frac{2c_{\omega}^{ts}g_1\Gamma_{12}^2}{\pi(1-\Gamma_{12})\lep^2}\right)\sqrt{\frac{\pi\Gamma_2}{g_1\Gamma_{12}}}\lep
\end{equation} then $\Omega^{ts}=\beta_{ts}\sqrt{1-(\Omega^{ts})^2}$, which yields
\begin{equation}
	\label{om_ts}
	\Omega^{ts}=\frac{\beta_{ts}}{\sqrt{1+\beta_{ts}^2}}.
\end{equation}

We plot this form of $\Omega^{ts}$ as a function of $\Gamma_{12}$ in Fig. \ref{latt_num} where we have taken $\ep=0.0358$ and $\a_1=\a_2=1$ (note that this parameter set corresponds to set `ES3' in \cite{AM}). This provides good agreement with simulations of the full Gross Pitaevskii energy and confirms that our point energy (\ref{renorm}) well describes the system.

\section{Disk Plus Annulus}

Equation (\ref{spatsep}) introduces the critical $\bar{\a}_0$ (or equivalently $\Gamma_{12}$) for which the two disk state is no longer a solution. In this section we consider $\a_0>\bar{\a}_0$ so that the system is given by a disk in component-1 and an annulus in component-2. This requires $\a_1$ to be different to $\a_2$, because otherwise the conditions $\a_0>\bar{\a}_0$ and $\alpha_0^2-\alpha_1\alpha_2\leq 0$ are not consistent. Defining the inner and outer radii of component-2 to be at $r=R_{2_-}$ and $r=R_{2_+}$ such that $R_{2_-}<R_1<R_{2_+}$, the appropriate density profiles are
\begin{equation}
    \label{uh1}
    \rho_{TF,1}(r)=|\eta_1|^2=\frac{\mu_1-r^2}{2\a_1}
\end{equation}
for $0<r<R_{2_-}$,
\begin{eqnarray}
    \label{uh2}
    \rho_{TF,1}(r)=|\eta_1|^2&=&\frac{1}{2\a_1\Gamma_{12}}\left(\mu_1-\frac{\a_0}{\a_2}\mu_2-
    r^2\Gamma_2\right)\\
    \label{uh3}
    \rho_{TF,2}(r)=|\eta_2|^2&=&\frac{1}{2\a_2\Gamma_{12}}\left(\mu_2-\frac{\a_0}{\a_1}\mu_1
    -r^2\Gamma_1\right)
\end{eqnarray}
for $R_{2_-}<r< R_1$, and
\begin{equation}
    \rho_{TF,2}(r)=|\eta_2|^2=\frac{\mu_2-r^2}{2\a_2}
    \label{uh4}
\end{equation}
for $R_1<r< R_{2_+}$.

The normalisation condition (\ref{norm1}) gives the following expressions for the radii
\begin{eqnarray}
\label{eq:r22}
R_{2_-}&=&\sqrt{R-\sqrt{-\frac{\Gamma_2}{\Gamma_1}}S}\\
\label{eq:r11}
R_1&=&\sqrt{R-\sqrt{-\frac{\Gamma_1}{\Gamma_2}(1-\Gamma_2)^2}S}\\
\label{eq:r21}
R_{2_+}&=&\sqrt{R+\sqrt{-\Gamma_1\Gamma_2}S}
\end{eqnarray}
and
\begin{eqnarray}
\label{cm1}
\mu_1&=&R\\
\mu_2&=&R+\sqrt{-\Gamma_1\Gamma_2}S
%
% \mu_1&=&\left(\frac{4\tilde{\a}_1(1+\frac{\tilde{\a}_2^2}{\tilde{\a}_{0}^2}(1-\Gamma_2)^2)}
% {\pi}\right)^{1/2}\\
% \mu_2&=&\left(\frac{4\tilde{\a}_1(1+\frac{\tilde{\a}_2^2}{\tilde{\a}_{0}^2}(1-\Gamma_2)^2)}
% {\pi}\right)^{1/2}\\&&+\left(-\frac{4\Gamma_1\Gamma_2\tilde{\a}_1\tilde{\a}_2^2(1-\Gamma_2)}
% {\pi \tilde{\a}_{0}^2}\right)^{1/2}.
\label{cm2}
\end{eqnarray}
for the chemical potentials. Here we have introduced the parameters
\begin{eqnarray}
	R&=&2\sqrt{\frac{\tilde{\a}_1(1+\frac{\tilde{\a}_2^2}{\tilde{\a}_0^2}(1-\Gamma_2)^2)}
{\pi}}\\
	S&=&2\sqrt{\frac{\tilde{\a}_1\tilde{\a}_2^2(1-\Gamma_2)}
{\pi \tilde{\a}_{0}^2}}.
\end{eqnarray}

A number of conditions can immediately be found from these expressions. In order for $R_{2_-}<R_1$, it must be that $\Gamma_{12}>0$ [The inequalities $R_{2_+}>R_1$ and $R_{2_+}>R_{2_-}$ are automatically valid]. Secondly, all the expressions (except that for $\mu_1$) require $\Gamma_1\Gamma_2<0$.  This is equivalent to $\a_2>\a_1$ (under the assumption that the annulus develops in component-2). Thus the range of $\Gamma_{12}$ for which component-1 is circular and component-2 is annular  is given by $0<\Gamma_{12}<\bar{\Gamma}_{12}$, where $\bar{\Gamma}_{12}=\Gamma_{12}$ evaluated at $\a_{0}=\bar{\a}_{0}$ given by (\ref{spatsep}).

When $\a_{0}=\bar{\a}_{0}$ we have shown in Eq. (\ref{omc}) that the critical velocity for creation of the first vortex (that occurs at the origin in component-2) is identically zero. This is precisely the point at which the annulus develops ($\rho_{TF,2}(r=0)=0$ with $R_{2_-}=0$). In the region $\a_{0}>\bar{\a}_{0}$,  the first appearance of a topological defect is the development of a giant vortex in component-2 and this can be analyzed fully.
 We need to introduce
\begin{eqnarray}
	&\Lambda_2&=\frac{\a_2}{\tilde{\a_2}}\int_{R_{2_-}}^{R_{2_+}}\frac{\rho_{TF,2}(s)}{s}ds\nonumber\\
	&=&\frac{1}{4\tilde{\a_2}\Gamma_{12}}\left[\Gamma_1R_{2_-}^2\log\left({\frac{R_1^2}{R_{2_-}^2}}\right)
+\Gamma_{12}R_{2_+}^2\log\left({\frac{R_{2_+}^2}{R_1^2}}\right)\right].\label{lambda2}
\end{eqnarray} Then the critical velocity for nucleation of the giant vortex is determined by $\Omega_{gv}=\eps \Lambda_2$
 and the circulation of this giant vortex is the integer part of $\Omega/(\eps \Lambda_2)$. Next, we want to determine vortices in the bulk
  and apply (\ref{renomm}) to the case of the disk plus annulus.
 Thus we need to define  $X_2(r)$ to take into account the giant vortex.

 We define $X_1(r)$ the primitive of $-r\rho_{TF,1} (r)$ which vanishes at $R_1$ giving
\begin{equation}
	\begin{split}
	X_1(r)=&\frac{1}{8\a_1} (R_1^2-r^2)^2\\&+\frac{\a_0\Gamma_1}{8\a_1\a_2\Gamma_{12}}(R_{2_-}^2-R_1^2)(R_1^2+R_{2_-}^2-2r^2)
	\end{split}
\end{equation}
for $0<r<R_{2_-}$, and is the same as Eq. (\ref{X1}) in the region $R_{2_-}<r< R_1$. In order to account for the development of vortices beyond the
 giant vortex, we must recall that additional to the term $-\eps\Omega\int r \rho_{TF,2} (if_2,\nabla f_2)$, there will be a kinetic energy term
  coupling the giant vortex and the vortex cores, namely $\eps^2 d\int \rho_{TF,2}/r(if_2,\nabla f_2)$, where $d$ is the degree of the giant vortex,
   of order $\Omega/(\eps \Lambda_2)$. Therefore,
 we  define $X_2(r)$ to be the primitive of $-r\rho_{TF,2} (r)+(1/\Lambda_2)\rho_{TF,2} (r)/r $ which vanishes at both $R_{2_-}=0$ and $R_{2_+}=0$,
\begin{equation}
	X_2(r)=\int_r^{R_{2_+}}s\rho_{TF,2}(s)d{s}-\frac{1}{ \Lambda_2}\int_r^{R_{2_+}}\frac{\rho_{TF,2}(s)}{s}d{s}
\end{equation} where $\Lambda_2$ is given by (\ref{lambda2}).
Completing the integrals gives
\begin{equation}
	\begin{split} X_2(r)=&\frac{1}{8\a_2}(R_{2_+}^2-r^2)^2-\frac{\a_0\Gamma_2}{8\a_1\a_2\Gamma_{12}}(R_1^2-r^2)^2\\&-\frac{1}{4{\a}_2\Gamma_{12}\Lambda_2}\Bigg(\Gamma_1{R_{2_-}^2}\log\left(\frac{R_1^2}{r^2}\right)
	\\&+\Gamma_{12}{R_{2_+}^2}\log\left(\frac{R_{2_+}^2}{{R_1^2}}\right)+\Gamma_1 r^2-\Gamma_1 R_{2_-}^2\Bigg)
\end{split}
\end{equation}
for $R_{2_-}<r< R_1$ and
\begin{equation}
	\begin{split}
	X_2(r)=&\frac{1}{8\a_2}(R_{2_+}^2-r^2)^2\\&-\frac{1}{4{\a}_2\Lambda_2}\left({R_{2_+}^2\log\left(\frac{R_{2_+}^2}{r^2}\right)+r^2-R_{2_+}^2}\right)
\end{split}
\end{equation}
for $R_1<r<R_{2_+}$.
\begin{figure}
\begin{center}
\includegraphics[scale=0.4]{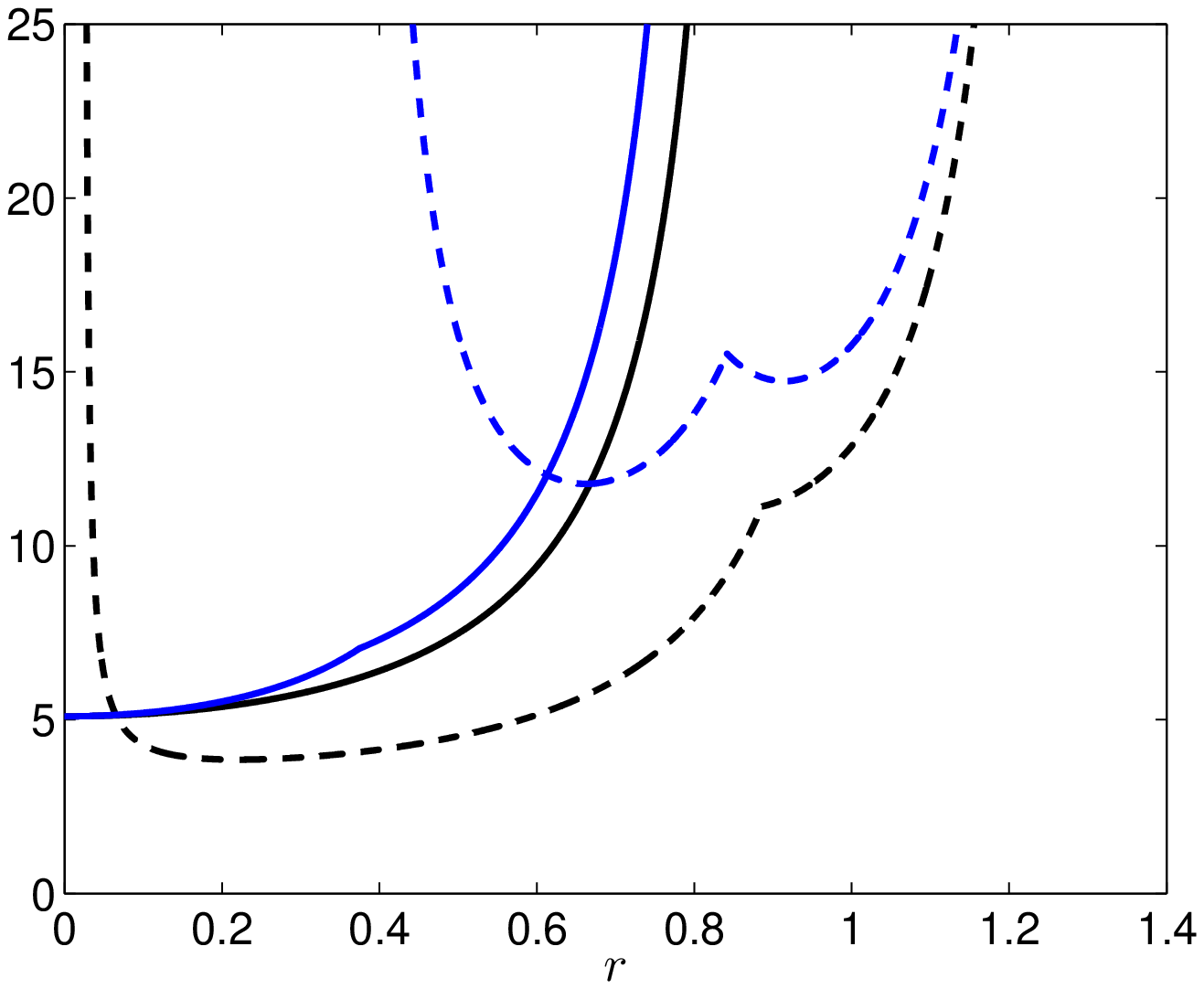}\\
\includegraphics[scale=0.4]{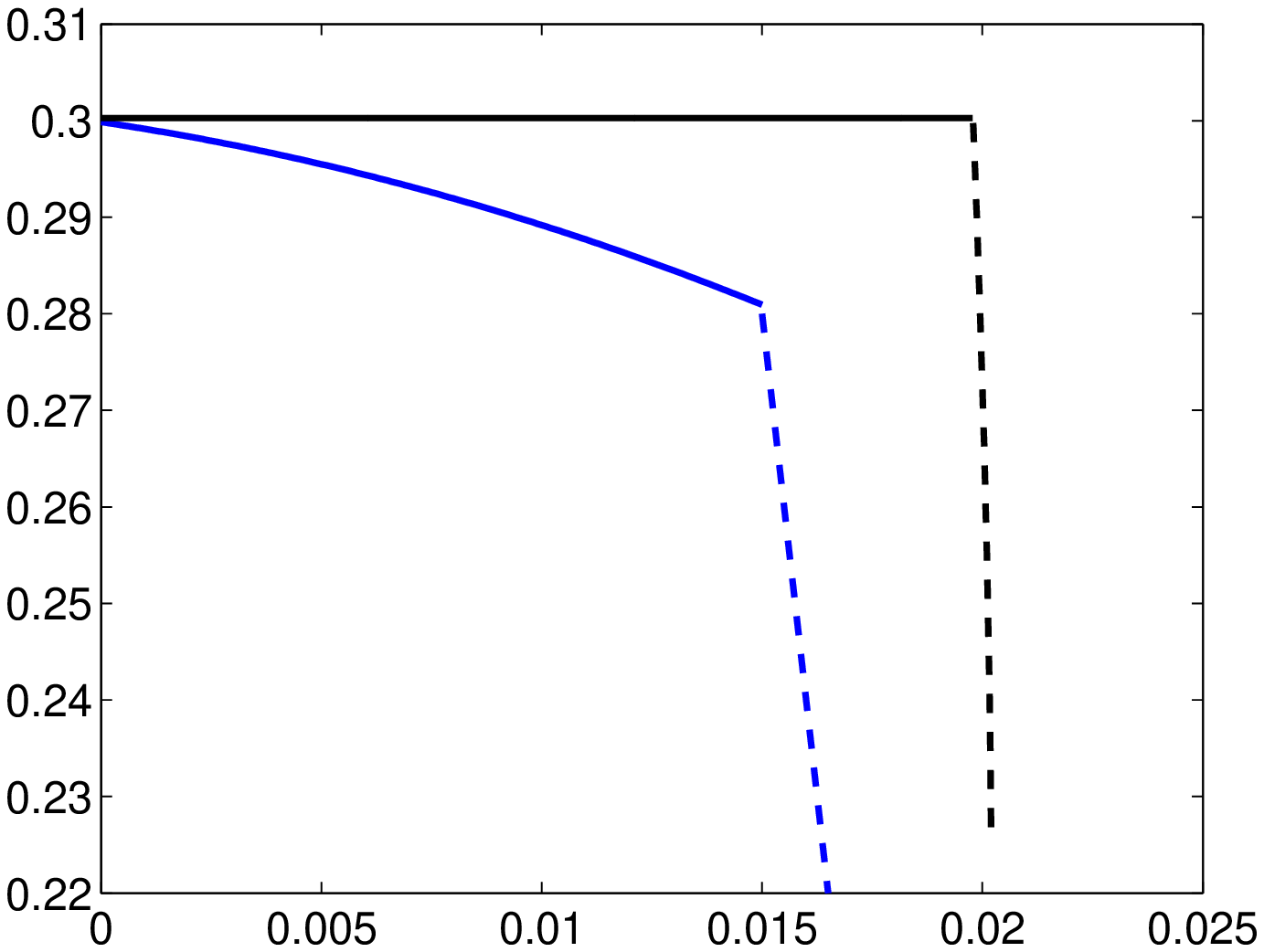}
\end{center}
\caption{(a) Functions ${\rho_{TF,1}}/{X_1}$ (solid lines) and ${\rho_{TF,2}}/{X_2}$ (dashed lines) given by Eq's (\ref{da1})-(\ref{da4}) for $\Gamma_{12}=0.0202$ (black lines) and $\Gamma_{12}=0.015$ (blue lines). (b) The critical velocity for creation of the first vortex plotted from Eq. (\ref{omegacc}) [black line] and numerically [blue line] as a function of $\Gamma_{12}$. The solid part corresponds to a vortex
 in component 1 at the center and the dashed part in component 2 in the interface region. The parameters are $\ep=0.0352$, $\a_1=0.97$ and $\a_2=1.03$. Note that $\bar\Gamma_{12}=0.02$}\label{da}
\end{figure}
As before we write down the ratios ${\rho_{TF,i}}/{X_i}$:
\begin{equation}
	\label{da1}
	\begin{split}
	 \frac{\rho_{TF,1}}{X_1}(r)=&\frac{4}{(R_1^2-r^2)}-\left({\frac{4\Gamma_1\a_0}{\Gamma_{12}\a_2}\frac{(R_{2_-}^2-R_1^2)(R_{2_-}^2-r^2)}{(R_1^2-r^2)}}\right)\\&\times\Big[(R_1^2-r^2)^2\\&+\frac{\Gamma_1\a_0}{\Gamma_{12}\a_2}(R_{2_-}^2-R_1^2)(R_1^2+R_{2_-}^2-2r^2)\Big]^{-1}
\end{split}
\end{equation}
for $0<r<R_{2_-}$,
\begin{eqnarray}
	\label{da2}
	\frac{\rho_{TF,1}}{X_1}(r)&=&\frac{4}{(R_1^2-r^2)},\\
	\label{da3}
	\frac{\rho_{TF,2}}{X_2}(r)&=&\frac{4\Gamma_1}{\Gamma_{12}}(R_{2_-}^2-r^2)\nonumber\\
	&&\quad\times\Bigg[(R_{2_+}^2-r^2)^2-\frac{\a_0\Gamma_2}{\a_1\Gamma_{12}}(R_1^2-r^2)^2\nonumber\\
	&&\qquad-\frac{2}{\Lambda_2\Gamma_{12}}\Bigg(\Gamma_1R_{2_-}^2\log\left(\frac{R_1^2}{r^2}\right)\nonumber\\
	&&\qquad+\Gamma_{12}R_{2_+}^2\log\left(\frac{R_{2_+}^2}{R_1^2}\right)+\Gamma_1(r^2-R_{2_-}^2)\Bigg)\Bigg]^{-1}
\end{eqnarray}
for $R_{2_-}<r< R_1$ and
\begin{equation}
	\label{da4}
	 \frac{\rho_{TF,2}}{X_2}(r)=\frac{4(R_{2_+}^2-r^2)}{(R_{2_+}^2-r^2)^2-\frac{2}{\Lambda_2}\left(R_{2_+}^2\log\left(\frac{R_{2_+}^2}{r^2}\right)+r^2-R_{2_+}^2\right)}
\end{equation}
for $R_1<r<R_{2_+}$.

It follows that
\begin{equation}
	\label{cas}
	\min\left(\frac{\rho_{TF,1}}{X_1}\right)=
	\begin{cases}
		4\frac{\a_2\Gamma_2R_1^2+\a_0\Gamma_1R_{2_-}^2}{\a_2\Gamma_2R_1^4+\a_0\Gamma_1R_{2_-}^4}\quad&\text{for}\qquad 0<r<R_{2_-}\\
		\frac{4}{(R_1^2-R_{2_-}^2)}\quad&\text{for}\qquad R_{2_-}<r< R_1.
	\end{cases}
\end{equation}
Using the expressions above for the radii, we see that
\begin{equation}
	\label{cas2}
	 4\frac{\a_2\Gamma_2R_1^2+\a_0\Gamma_1R_{2_-}^2}{\a_2\Gamma_2R_1^4+\a_0\Gamma_1R_{2_-}^4}=\frac{2}{\tilde{\a_1}}\sqrt{\frac{\pi}{\a_2}\left(\a_1\tilde{\a}_2+\tilde{\a}_1\a_2\right)}
\end{equation}
which we note is independent of $\a_0$. If we compare the values of Eq. (\ref{cas}), we see that they are equal when $\a_0=\bar{\a}_0$ (which we can consider a degenerate case) and otherwise $\min({\rho_{TF,1}}/{X_1})$ over the whole space is simply given by the expression in Eq. (\ref{cas2}). This implies that the vortex in component-1 is preferred at the origin rather than in the interface region of the two components.

To complete the analysis we must also find the $\min({\rho_{TF,2}}/{X_2})$ over the whole space to determine whether vortices
 appear first in component 1 or component 2. It is not possible to find an analytic expression for this minimum and so we must resort to plotting the function and comparing this to $\min({\rho_{TF,1}}/{X_1})$. In Fig. \ref{da}, we plot the functions ${\rho_{TF,1}}/{X_1}$ and ${\rho_{TF,2}}/{X_2}$
in two different cases where $\min({\rho_{TF,1}}/{X_1})<\min({\rho_{TF,2}}/{X_2})$ and $\min({\rho_{TF,1}}/{X_1})>\min({\rho_{TF,2}}/{X_2})$. We also plot the critical $\Omega$ defined in Eq. (\ref{omegacc}) as a function of $\Gamma_{12}$. We see that until some critical value of $\Gamma_{12}$,
  the vortex first nucleates at the origin in component-1, while above in some small region in $\Gamma_{12}$ before reaching $\bar \Gamma_{12}$,
   the vortex first nucleates in the interface region in the second component. Then, increasing $\Omega$, leads to a vortex lattice close to this point of nucleation. Recall that the giant vortex is always present.

\section*{Conclusion} From the energy of a rotating two component condensate, we have derived a
 reduced energy (\ref{renorm}) governing the location of peaks and vortices. We have found that the
  ground state of this reduced energy yields a square lattice of vortices in regimes consistent with the
  ones found numerically for the full Gross Pitaevskii energy. We have analyzed in detail the geometry
  of the ground state (two disks or disk and annulus) and derived formula of the critical velocity for the appearance
  of the first vortex. We can determine in which component the first vortex appears and the shape of the lattice.

\section*{Acknowledgments}

The authors  acknowledge
support from the French ministry Grant ANR-BLAN-0238, VoLQuan.

\end{document}